\def\tr{\hbox{tr}}
\documentstyle[aps,psfig]{revtex}

\begin{document}

\title{A further study of the possible scaling region of lattice chiral fermions}
\author{She-Sheng Xue}
\address{I.C.R.A.-International Center for Relativistic Astrophysics and
Physics Department, University of Rome ``La Sapienza", 00185 Rome, Italy}

\maketitle

\begin{abstract}

In the possible scaling region for an $SU(2)$ lattice chiral fermion advocated in {\it Nucl.~Phys.} B486 (1997) 282, no hard spontaneous symmetry breaking occurs and doublers are gauge-invariantly decoupled via mixing with composite three-fermion-states that are formed by local multifermion interactions. However the strong coupling expansion breaks down due to
no ``static limit'' for the low-energy limit ($pa\sim 0$). In both neutral and charged channels, we further analyze relevant truncated Green functions of three-fermion-operators by the strong coupling expansion and analytical continuation of these Green functions in the momentum space. It is shown that in the low-energy limit, these relevant truncated Green functions of three-fermion-states with the ``wrong'' chiralities positively vanish due to the generalized form factors (the wave-function renormalizations) of these composite three-fermion-states vanishing as $O((pa)^4)$ for $pa\sim 0$. This strongly implies that the composite three-fermion-states with ``wrong''
chirality are ``decoupled'' in this limit and the low-energy spectrum is chiral, as a consequence, chiral gauge symmetries can be exactly preserved.
 
\end{abstract}
 
\vglue 1cm

\section{Introduction}

The attempt to get around the ``no-go'' theorem \cite{nn81} for the
``vector-like'' phenomenon of chiral fermions on a lattice is currently a very
important issue of theoretical particle physics(\cite{review}--\cite{tdlee}).
One attempt is the approach of multifermion couplings, which can be traced
back from the recent lattice formulation of the Standard Model\cite{mc} to the
pioneer model suggested by Eichten and Preskill in ref.\cite{ep} and successive
work\cite{xueall,xue97,xue97l} in the past years. However, on the other hand, it was
pointed out in an crucial paper\cite{gpr} that the models proposed in
ref.\cite{ep} fail to give chiral fermions in the continuum limit. The reasons
are that a Nambu-Jona
Lasinio (NJL)\cite{njl} spontaneous symmetry breaking phase separates the
strong-coupling symmetric phase from the weak-coupling symmetric phase, and the
right-handed Weyl states do not completely disassociate from the left-handed
chiral fermions. Further, another crucial paper\cite{sy93} tried to extent the
``no-go'' theorem into interacting theories based on the plausible argument of
such theories being local. The definite
failure of the models so constructed has then been a general belief\cite{review}
and it has been very doubtful that exact chiral gauge symmetries could be realized in a lattice. 
Nevertheless, based on the consistency of the Rome-approach\cite{rome} to an $SU(2)-$chiral gauge theory on the lattice and physical nature of chiralities and the cutoff(lattice), we did strongly believe the existence of exact chiral gauge symmteries on the lattice and have been working on the multifermion couplings models for for many years\cite{xueall,xue97,xue97l}. 

For reader's convenience, we briefly describe the crucial points of 
the Eichten and Preskill \cite{ep}
idea follows. Multifermion couplings are
introduced such that, in the phase space of strong-couplings, Weyl states
composing three elementary Weyl fermions (three-fermion states) are bound.
Then, these three-fermion states pair up with elementary Weyl fermions to be
Dirac fermions. Such Dirac fermions can be massive without violating chiral
symmetries due to the appropriate quantum numbers and chirality carried by
these three-fermion states. The binding thresholds of such three-fermion states
depend on elementary Weyl modes residing in different regions of the Brillouin
zone. If one assumes that the spontaneous symmetry breaking of the NJL-type does not occur and such binding thresholds separate the weak
coupling symmetric phase from the strong-coupling symmetric phase, there are
two possibilities to realize the continuum limit of chiral fermions in phase
space. One is of crossing over the binding threshold the three-fermion
state of chiral fermions. Another is of a wedge between two thresholds, where
the three-fermion state of chiral fermions has not been formed, provided all
doublers sitting in various edges of the Brillouin zone have been bound to be
massive Dirac fermions and decouple.

We should not be surprised that a particular model of multifermion couplings
does not work. This does not means that Eichten and Preskill's idea is definitely wrong unless there
is another generalized ``no-go'' theorem on interacting theories \cite{sy93} for
a whole range of coupling strength. Actually, Nielsen and Ninomiya gave an interesting 
comment on Eichten and Preskill's idea
based on their intuition of anomalies \cite{nn91}. To conclude, we believe that further considerations
of constructing chiral fermions on the lattice with multifermion couplings and
careful studies of the spectrum in each phase of theories so constructed are
necessary. In the previous paper \cite{xue97l} we illustrated a possible scaling 
regin for lattice chiral fermions in the phase space of multifermion couplings. 
In this paper, as a successive paper of \cite{xue97l} we go further to demonstrate
the existence of the possible scaling regin for lattice chiral fermions by computing
relevant Green functions and the analytical property of these Green functions in the momentum space.

The paper is organized as follow: section II summarizes the main points and results of the
possible scaling region found in
the previous paper \cite{xue97l}; in section III we discuss the momentum-dependent static limit and its analytical property in the momentum space, which is a crucial feature to have 
a possible scaling regin for lattice chiral fermions; in section IV we discuss the notion 
of three-fermion-cuts in the possible scaling regin; in section V we present very solid computations of relevant Green functions relating to
the three-fermion-cuts for the neutral channel and conclude that neutral chiral fermions do exist in the possible scaling regin, whic already violates the ``no-go'' theorem; in section VI based on 
approximate computations of relevant Green-functions relating to the three-fermion-cuts for the charged channel and the analytical property of these Green-function in the momentum space, we show charged lattice chiral fermions could exist in such a scaling regin; in section VII we discuss the crucial properties of our model to work, what and why our model are different from  the Eichten-Preskill and Smit-Swift model\cite{ss}. In the final section, we make some remarks and outlook regarding this approach to lattice chiral gauge theories.

In this paper, we are not pretending to solve all problems of
lattice chiral gauge theories in this paper. If chiral fermions are gauged, the
important questions concerning correct and consistent features of gauge
bosons and the coupling between gauge bosons and chiral fermions are open,
but beyond the scope of this paper. Other important questions \cite{anomaly} concerning
about gauge anomalies and anomalous global currents (instanton effects and
non-conservation of fermion currents), which Eichten and Preskill \cite{ep} suggested to obtain
by explicitly breaking the global symmetries associating to these currents,
will be studied in separate papers.

\section{A possible scaling region of lattice chiral fermions}

To more easily show the possibility and reason of dynamics for such 
constructed models to 
work, we further analyze the simple and anomaly-free model of multifermion
couplings proposed in the ref.\cite{xue97}.
Note that $\psi^i_L$ ($i=1,2$) is an $SU_L(2)$ gauged
doublet, $\chi_R$ is an $SU_L(2)$ singlet and both are two-component Weyl
fermions. $\chi_R$ is treated as a ``spectator'' fermion. 
$\psi^i_L$ and $\chi_R$ fields are dimensionful $[a^{{1\over2}}]$. 
The following action for chiral fermions 
with the $SU_L(2)\otimes U_R(1)$ chiral symmetries on the lattice is suggested:
\begin{eqnarray}
S&=&S_f+S_1+S_2,\label{action}\\
S_f&=&{1\over 2a}\sum_x\sum_\mu\left(\bar\psi^i_L(x)\gamma_\mu D^\mu_{ij}\psi^j_L(x)+
\bar\chi_R(x)\gamma_\mu\partial^\mu\chi_R(x)\right),\nonumber\\
S_1&=&g_1
\sum_x\bar\psi^i_L(x)\cdot\chi_R(x)\bar\chi_R(x)\cdot\psi_L^i(x),
\nonumber\\
S_2&=&g_2\sum_x \bar\psi^i_L(x)\cdot\left[\Delta\chi_R(x)\right]
\left[\Delta\bar\chi_R(x)\right]\cdot\psi_L^i(x),\nonumber
\end{eqnarray}
where $S_f$ is the naive lattice action for chiral fermions, 
$a$ is the lattice spacing.
$S_1$ and $S_2$ are two external multifermion 
couplings, where the $g_1$ and $g_2$ have dimension
$[a^{-2}]$, and the Wilson factor\cite{wilson} is given as,
\begin{eqnarray}
\Delta\chi_R(x)&\equiv&\sum_\mu
\left[ \chi_R(x+\mu)+\chi_R(x-\mu)-2\chi_R(x)\right],\nonumber\\
2w(p)&=&\int_xe^{-ipx}\Delta(x)=\sum_\mu\left(1-\cos(p_\mu)\right).
\label{wisf}
\end{eqnarray}
Note that all momenta are scaled to be
dimensionless, $p=\tilde p+\pi_A$
where $\pi_A$ runs over fifteen lattice momenta ($\pi_A\not=0$).

The action (\ref{action}) has an exact local $SU_L(2)$ chiral gauge symmetry,
\begin{equation} 
\sum_\mu \gamma_\mu D^\mu=\sum_\mu(U_\mu(x)\delta_{x,x+\mu}
-U^\dagger_\mu (x)\delta_{x,x-\mu}),\hskip0.5cm U_\mu(x)\in SU_L(2),
\label{kinetic} 
\end{equation} 
which is the gauge symmetry that the continuum theory
(the target theory) possesses. The global flavour symmetry $U_L(1)\otimes
U_R(1)$ is not explicitly broken in eq.~(\ref{action}),
we will not discuss the property of violating fermion number in such a  
model.

It has been advocated\cite{xue97} there exists a plausible scaling region, which
is a peculiar segment in the phase space of the multifermion couplings $g_1,g_2$, 
\begin{equation}
{\cal A}=\Big[g_1\rightarrow 0, g_2^{c,a}<g_2<g_2^{c,\infty}\Big],\hskip0.3cm
a^2g_2^{c,a}=0.124,\hskip0.3cm
1\ll g_2^{c,\infty}< \infty, 
\label{segment}
\end{equation}
for chiral fermions in the low-energy limit. $g_2^{c,\infty}$  is a finite 
number and $g_2^{c,a}$ indicates the
critical value above which the effective multifermion couplings associating to all doublers are
strong enough, so that all doublers are gauge-invariantly decoupled. We
qualitatively determined $a^2g_2^{c,a}=0.124$ in ref.\cite{xue97}. The crucial
points for this scaling region to exist are briefly described in the following.

In segment ${\cal A}$ (\ref{segment}), the
action (\ref{action}) possesses a $\chi_R$-shift-symmetry \cite{gp},
i.e., the action is invariant under the transformation:
\begin{equation}
\bar\chi_R(x) \rightarrow \bar\chi_R(x)+\bar\epsilon,\hskip1.5cm
\chi_R(x) \rightarrow \chi_R(x)+\epsilon,
\label{shift}
\end{equation}
where $\epsilon$ is independent of space-time.
The Ward identity corresponding to this
$\chi_R$-shift-symmetry is given as\cite{xue97}
($g_1\rightarrow 0$),
\begin{equation}
{1\over 2a}\gamma_\mu\partial^\mu\chi'_R(x)
+g_2\!\langle\Delta\!\left(\bar\psi^i_L(x)\!\cdot\!\left[
\Delta\chi_R(x)\right]\psi_L^i(x)\right)\rangle-{\delta\Gamma\over\delta\bar
\chi'_R(x)}=0,
\label{w}
\end{equation}
where the ``primed'' fields are defined through the generating functional
approach, and ``$\Gamma$'' is the effective potential. The important consequences 
of this Ward identity in segment ${\cal A}$ are:
\begin{itemize}
\item the low-energy mode ($p\sim 0$) of $\chi_R$ is a free mode and decoupled:
\begin{equation}
\int_x e^{-ipx} {\delta^{(2)}\Gamma\over\delta\chi'_R(x)\delta\bar\chi'_R(0)}
={i\over a}\gamma_\mu\sin(p^\mu);
\label{free}
\end{equation}

\item no hard spontaneous chiral symmetry breaking ($O({1\over a})$) occurs
(see eqs.(30) and (31) in ref.\cite{xue97})\footnote{The soft symmetry 
breaking for the low-energy modes ($p\sim 0$) is allowed.},
\begin{equation}
\int_x e^{-ipx} {\delta^{(2)}\Gamma\over\delta\psi'^i_L(x)\delta\bar\chi'_R(0)}
={1\over2}\Sigma^i(p)=0 \hskip0.5cm p=0,
\label{ws2}
\end{equation}
in addition, we have:
\begin{equation}
\Sigma(p)=0\hskip0.5cm p\not=0
\label{ws2'}
\end{equation}
which is shown by the strong coupling expansion (see eq.(104) in \cite{xue97}).

\end{itemize}

For the strong coupling $g_2\gg 1$ in the segment ${\cal A}$, the following 
three-fermion-states
comprising of the elementary fields $\psi^i_L$ and $\chi_R$ in (\ref{action}) are bound:
\begin{equation}
\Psi_R^i={1\over
2a}(\bar\chi_R\cdot\psi^i_L)\chi_R;\hskip1cm\Psi^n_L={1\over 2a}(\bar\psi_L^i
\cdot\chi_R)\psi_L^i.
\label{bound}
\end{equation}
These fermion bound states possess the ``wrong'' chiralities in  contrast with 
the ``right'' chiralities possessed by the elementary fields $\psi^i_L$ and $\chi_R$.
The two-point Green functions of these charged $\Psi^i_R$ and 
neutral $\Psi^n_L$ have poles at the total momentum $p=\pi_A$\cite{xue97},
\begin{eqnarray}
S^{ij}_{MM}(p)&=&\int_x e^{-ipx}\langle\Psi^i_R(0)\bar\Psi^j_R(x)\rangle
\simeq\delta_{ij}{{i\over a}\sum_\mu\sin p^\mu\gamma_\mu\over
{1\over a^2}\sum_\mu\sin^2 p_\mu+M^2(p)}P_L;
\label{sc11}\\
S^n_{MM}(p)&=&\int_xe^{-ipx} \langle\Psi^n_L(0)\bar\Psi^n_L(x)\rangle
\simeq {{i\over a}\sum_\mu\sin p^\mu\gamma_\mu\over
{1\over a^2}\sum_\mu\sin^2 p_\mu+M^2(p)}P_R,
\label{sn11}\\
M(p)&=&8ag_2w^2(p),
\label{m}
\end{eqnarray}
where $p\sim\pi_A$ and $w^2(p)\not=0$. And the two-point Green functions 
for doublers ($p\sim\pi_A$) of the elementary fields $\chi_R$ and $\psi_L^i$ are 
given by,
\begin{eqnarray}
S_{LL}^{ij}(p)=\int_xe^{-ipx}\langle\psi_L^i(0)\bar\psi^j_L(x)\rangle &\simeq&
\delta_{ij}{{i\over a}\gamma_\mu \sin (p)^\mu
\over {1\over a^2}\sin^2p + M^2(p)}P_R;\label{lp'}\\
S_{RR}(p)=\int_xe^{-ipx}\langle\chi_R(0)\bar\chi_R(x)\rangle 
&\simeq&{{i\over a}\gamma_\mu \sin (p)^\mu 
\over {1\over a^2}\sin^2p + M^2(p)}P_L.
\label{rp'}
\end{eqnarray}
The three-fermion-states (\ref{bound},\ref{sc11},\ref{sn11}) are Weyl
fermions and respectively mix with the doublers of the 
elementary Weyl fields $\bar\chi_R$ and $\bar\psi_L^i$ (\ref{lp'},\ref{rp'}),
\begin{eqnarray}
S^{ij}_{ML}(p)
=\int_xe^{-ipx}\langle\Psi_R^i(0)\bar\psi_L^j(x)\rangle &\simeq& \delta_{ij}{M(p)
\over {1\over a^2}\sin^2p + M^2(p)}P_R,
\label{mixingc}\\
S^n_{MR}(p)=\int_xe^{-ipx}\langle \Psi^{n}_L(0)\bar\chi_R(x)\rangle &\simeq&
{M(p)\over {1\over a^2}\sin^2p + M^2(p)}P_L.
\label{mx}
\end{eqnarray}
As a result, the neutral $\Psi_n$ and 
charged 
$\Psi_c^i$ Dirac fermions are formed\footnote{The propagators of these
Dirac fermions can be obtained by summing all relevant Green functions shown
in above.}, 
\begin{equation}
\Psi^i_c=(\psi_L^i, \Psi^i_R);\hskip1cm\Psi_n=(\Psi_L^n, \chi_R),
\label{di}
\end{equation}
and the spectrum is vector-like and massive.
These show that all doublers
are decoupled as very massive Dirac fermions consistently with the 
$SU_L(2)\otimes U_R(1)$ chiral symmetries, 
since the three-fermion-states (\ref{bound}) carry the appropriate quantum
numbers of the chiral groups that accommodate $\psi^i_L$ and
$\chi_R$. 

\section{The momentum-dependent static limit and its analytical property}

Eqs.(\ref{sc11},\ref{sn11}) for doublers is obtained by the the strong-coupling 
expansion in powers of ${1\over g_2}$. 
For the strong coupling $g_2\gg 1$, the kinetic terms 
can be dropped and the strong-coupling limit is given as,
\begin{eqnarray} 
Z&=&\Pi_{xi\alpha}\int[d\bar\chi_R^\alpha (x) d\chi_R^\alpha (x)]
[d\bar\psi_L^{i\alpha}(x) d\psi_L^{i\alpha}(x)]\exp\left(-S_2(x)\right)
\nonumber\\
&=&(2g_2)^{4N}\left(\det\Delta^2(x)\right)^4,\hskip0.3cm g_2\gg 1,
\label{stronglimit}
\end{eqnarray}
where the determent is taken only over the lattice space-time and $N$
is the number of lattice sites. For the non-zero eigenvalues of the operator
$\Delta^2(x)$ (\ref{wisf}), which are associating to the doublers ($p\simeq\pi_A$) of
$\psi^i_L(x)$ and $\chi_R(x)$, eq.~(\ref{stronglimit}) 
shows the existence of a sensible strong-coupling limit. Note
that the operator $\Delta(x)\sim 2w(p)$ (\ref{wisf},\ref{stronglimit}) 
has different eigenvalues $4\sim 6$ with respect to different
doublers $(p\sim\pi_A)$, and the strong
coupling expansion is actually in terms of powers of ${1\over 4g_2w^2(p)}$.
As the consequence, the two-point Green functions (\ref{sc11}-\ref{mx}) 
computed by the strong coupling expansion with respect to doublers should be 
a good approximation even for the intermediate coupling
\begin{equation}
a^2g_2\sim O(1).
\label{o(1)}
\end{equation}
This discussion agrees with the qualitatively determined critical value 
$g_2^{c,a}=0.124$ in (\ref{segment}), above which all doublers are decoupled 
via eqs.(\ref{di}).

However, as for the zero eigenvalues of the operator $\Delta^2(x)$, which
precisely correspond to the low-energy modes ($p\sim 0$) of $\psi^i_L(x)$ and
$\chi_R(x)$, this strong-coupling limit is trivial and the strong-coupling
expansion in powers of ${1\over g_2}$ breaks down. This physically means the
weakness of the effective multifermion coupling for such low-energy modes of
$\psi^i_L(x)$ and $\chi_R(x)$. Nevertheless, we cannot exclude the possibility
of low-energy modes of three-fermion-states (\ref{bound}), which are
represented by the poles $p\sim 0$ of the propagators (\ref{sc11},\ref{sn11}). 

On the other hand, as a consequence of the multifermion interacting action 
(\ref{action}) being local, in the strong coupling limit the effective action 
(inverse propagator),
which is bilinear in terms of interpolating fields, should be local and
analytical in the whole Brillouin zone. Thus, 
the ``no-go'' theorem of Nielsen and Ninomiya is still applicable to this
case \cite{sy93}. Based on such an observation, one might argue the existence of
the massless
spectrum of the charged and neutral 
three-fermion-states (\ref{bound}) by the analytic
continuation of their propagators (\ref{sc11},\ref{sn11})  
from $p\sim\pi_A$ to $p\sim 0$. As a result, the low-energy spectrum (\ref{di})
is also vector-like. 

Indeed, due to the locality of action (\ref{action}) presented in this paper,
all Green functions must be analytically continuous functions in 
energy-momentum space, provided the dynamics is fixed by given $g_1$ and $g_2$.
In the strong coupling symmetric phase (PSM) where $g_1\gg 1$ and a sensible
strong-coupling limit exists, two-point Green functions for
three-fermion-states (\ref{bound}) are essentially indistinguishable from that 
of the elementary fermion fields $\psi^i_L(x)$ and $\chi_R(x)$ appearing in the Lagrangian. In 
such a phase, the
analytical continuation of Green functions for both elementary and composite
fields in the whole momentum space does result in the ``vector-like'' 
phenomenon, as
asserted by the ``no-go'' theorem for a free-fermion theory. The only loophole
would appear if the propagators of interpolating fields (three-fermion-states)
properly vanished and no longer had poles at $p\sim 0$. This indicates that at
$p\sim 0$, these three-fermion-states dissolve into three-fermion-cuts, where the
``no-go'' theorem is entirely inapplicable. We discuss the concept of three-fermion-cuts
in the next section.

\section{Generalized form-factors of three-fermion-states}

We turn to discuss how these three-fermion-states (\ref{sn11},\ref{sc11}) 
with the ``wrong'' chiralities dissolve
into three-fermion-cuts in segment ${\cal A}$ for the low-energy limit 
($p\rightarrow 0$). These three-fermion-cuts\cite{xue97l,three}:
\begin{equation}
{\cal C}[\Psi_L^n(x)],\hskip1cm {\cal C}[\Psi_R^i(x)],
\label{cut}
\end{equation}
are the virtual states of three 
individual chiral fermions with a continuous energy spectrum, provided the 
total momentum $p$ is fixed. However, 
these virtual
states carry exactly the same quantum numbers and total momentum 
$p$ as that
of three-fermion-states. Thus, gauge symmetries are preserved in such a 
phenomenon of dissolving.
The dynamics of the three-fermion-states dissolving 
into their virtual state is that the negative binding energy
of three-fermion-states goes to zero. In the energy plane, it was shown that due
to the variety of effective interactions (potential), the poles for bound states
can be analytically continued to the cuts for virtual states on the physical
sheet\cite{smatrix}. Presumably, in this analytical property of effective 
interactions, no other dynamics, e.g., spontaneous symmetry breaking, takes 
place. In segment ${\cal A}$,
the weakness of effective multifermion couplings for the low-energy 
modes of $\psi_L^i$ and $\chi_R$ could lead to the vanishing of the binding energy 
of the three-fermion-states (\ref{bound}). We can conceive a ``dissolving'' scale
(threshold) $\epsilon$
\begin{equation}
\tilde v\ll\epsilon<{1\over a},
\label{disscale}
\end{equation}
where $\tilde v$ is the possible soft spontaneous symmetry breaking scale
($a\tilde v\simeq 0$).  
From inequality (\ref{disscale}), we understand that no hard
spontaneous symmetry breaking in segment ${\cal A}$ is extremely crucial
for the possibility of analytical continuation of propagators from poles for 
three-fermion-states to cuts for
virtual states of three individual fermions.\footnote{ At the dissolving scale 
$\epsilon\gg \tilde v$, we can approximately treat elementary massive fermions (e.g.~
massive particles in the Standard Model)  
as massless particles.} 

In the relativistic Lagrangian approach, to discuss the property of 
three-fermion-states dissolving into three-fermion-cuts, we are bound to 
dynamically calculate two-point functions of three-fermion-states (Fig.1)
to identify not only their poles, but also the corresponding residues. Using the
strong coupling expansion, we approximately determined the simple poles 
for doublers ($p\sim\pi_A$) in eqs.(\ref{sc11},\ref{sn11}). The residues 
$Z_R(p)$ and $Z_L(p)$ of these simple poles are defined as,
\begin{eqnarray}
S^{ij}_{MM}(p)&=&\int_x e^{-ipx}\langle\Psi^i_R(0)\bar\Psi^j_R(x)\rangle
\simeq\delta_{ij}{Z_R(p){i\over a}\sum_\mu\sin p^\mu\gamma_\mu Z_R(p)\over
{1\over a^2}\sum_\mu\sin^2 p_\mu+M^2(p)}P_L;
\label{rsc11}\\
S^n_{MM}(p)&=&\int_x e^{-ipx} \langle\Psi^n_L(0)\bar\Psi^n_L(x)\rangle
\simeq {Z_L(p){i\over a}\sum_\mu\sin p^\mu\gamma_\mu Z_L(p)\over
{1\over a^2}\sum_\mu\sin^2 p_\mu+M^2(p)}P_R.
\label{rsn11}
\end{eqnarray}
In fact, these residues represent the generalized form factors of three-fermion-states. 
The $Z_{L,R}(p)$ momentum dependence indicates that different doublers 
have different form factors, which implies the ``size'' of bound states is 
different from one doubler to another. This is clearly attributed to the momentum
dependence of effective multifermion couplings in action (\ref{action}). If 
these residues $Z_{L,R}(p=\pi_A)$ are positive\footnote{We do not want to have
ghost states with negative norm.} and finite constants with respect to each 
doubler, we can just make a wave-function renormalization of 
three-fermion-states with respect to each doubler,
\begin{equation}
\Psi_R^i|_{ren}=Z^{-1}_R\Psi_R^i;\hskip1cm\Psi^n_L|_{ren}=Z^{-1}_L\Psi^n_L,
\label{rbound}
\end{equation}
and the two-point Green functions (\ref{rsc11},\ref{rsn11}) turn into 
eqs.(\ref{sc11},\ref{sn11}) in terms of the renormalized fields (\ref{rbound}). 

The residues (generalized
form factors) $Z_{R,L}(p)$ (\ref{rsc11},\ref{rsn11}) of the three-fermion-states 
(\ref{bound}) are given by one-particle irreducible (1PI) truncated Green 
functions (see Fig.2),
\begin{equation}
Z_L(p)=\int_xe^{-ipx}
{\delta^{(2)}\Gamma\over\delta\Psi'^n_L(x)\delta\bar\chi'_R(0)},
\hskip1cm Z_R(p)=\int_xe^{-ipx}
{\delta^{(2)}\Gamma\over\delta\Psi'^i_R(x)\delta\bar\psi'^j_L(0)}.
\label{zlzr}
\end{equation}
The ``primed fields'' $\Psi'^n_R(x)$ and $\Psi'^i_R(x)$ of three-fermion-states
are defined by eqs.(41) and (42) in ref.\cite{xue97} through the generating
functional approach.
These generalized form factors $Z_L(p)$ and $Z_R(p)$  give the overlap
between three-fermion-operators
($\Psi^i_R(x)$, $\Psi^n_L(x)$) and the interpolating three-fermion-states,
that appear in the space of asymptotic states of the theory in the scaling
region.

This description coincides with the renormalization of n-point
1PI functions with insertions of composite operators. In general,
the renormalized n-point 1PI functions $\Gamma_{ren}^{(n)}$ with single and two
insertions of composite operators are given by\cite{itz}, 
\begin{eqnarray}
\Gamma_{ren}^{(n)}(p_1,q_1,q_2,\cdot\cdot\cdot,q_n)&=&
Z\Gamma_{reg}^{(n)}(p_1,q_1,q_2,\cdot\cdot\cdot,q_n),\nonumber\\
\Gamma_{ren}^{(n)}(p_1,p_2,q_1,q_2,\cdot\cdot\cdot,q_n)&=&
Z^2\Gamma_{reg}^{(n)}(p_1,p_2,q_1,q_2,\cdot\cdot\cdot,q_n),
\label{ren}
\end{eqnarray}
where $\Gamma_{reg}^{(n)}$ are the regularized n-point 1PI functions and $p_1$
and $p_2$ stand for the momenta entering the composite operators. 
Similarly given by eq.(\ref{zlzr}) (Fig.2)\cite{itz}, the $Z$'s
are the generalized  ``wave-function renormalizations'' of composite operators.
It is worthwhile to emphasize that for residues $Z_{L,R}$ being
positively finite, the wave-function renormalization of composite fields 
is the exactly same as the wave-function renormalization of elementary fields.
In fact, composite particles are indistinguishable
from elementary particles in this case. However, the
normal wave-function renormalizations of elementary fields appearing in the 
Lagrangian is attributed to the fact that these elementary fields are defined at 
different scales rather than their ``form factor''. Note that the elementary 
fields $\psi^i_L(x)$ and $\chi_R(x)$ in the action (\ref{action}) are bare 
fields and have not yet been renormalized. 

However, in the analytical continuation of the propagators 
(\ref{rsc11},\ref{rsn11}) in momentum space, let us assume an interesting case that the 
residues $Z_{L,R}(p)$ positively
vanish in the limit $p\rightarrow 0$ for the pole $p\sim 0$ in eqs.(\ref{rsc11},
\ref{rsn11}),
\begin{equation}
Z_{L,R}(p)\rightarrow O(p^n)\hskip0.5cm p\rightarrow 0,
\label{p=0}
\end{equation}
with $n=2,4,6,\cdot\cdot\cdot$. This property of $Z_{L,R}(p)$ 
could be realized by 
an appropriate momentum-dependence of the effective multifermion couplings 
$g_1,g_2$. Eq.(\ref{p=0}) implies that $p\sim 0$ is no longer a pole for a 
relativistic particle in the propagators (\ref{rsc11},
\ref{rsn11}). We are not allowed to make wave-function renormalization
(\ref{rbound}) with respect to $p\sim 0$. Eq.(\ref{p=0}) indicates that the 
``size'' of bound states (\ref{bound}) increases as $p\rightarrow 0$. 
These three-fermion-states may 
eventually dissolve into the virtual states of three individual fermions, whose
possible configuration in momentum space is $(p,p,-p)$\cite{xue97l,nn91}, 
where $p$ is the total momentum and the relative momentum ($q$) is zero. 
This dissolving phenomenon is entirely determined by both the dynamical and 
kinetic properties of the interacting theory.

This is reminiscent of the papers\cite{wein} discussing whether helium is
an elementary or composite particle based on the vanishing of wave-function 
renormalizations of composite states. It is normally referred to as the composite condition
that the wave-function renormalizations of bound states 
go to zero ($Z\rightarrow 0$)\cite{com}. So far, we only give an intuitive and 
qualitative discussion of the dissolving phenomenon on the basis of the 
relations between the residues (generalized form factors) $Z_{L,R}(p)$, 
renormalized three-fermion-states
and virtual states of three individual fermions (three-fermion-cut). Evidently, we are bound to 
do some dynamical calculations to show this phenomenon could happen.

\section{The three-fermion-cut for neutral channel}

The form factor $Z_L(p)$ in eq.(\ref{zlzr}) for the left-handed
three-fermion-state $\Psi_L^n$ can be completely determined by the Ward identity (\ref{w}). We
appropriately take functional derivative of the Ward identity (\ref{w}) with
respect to the ``primed'' field $\Psi'^n_L(x)$, and we obtain,
\begin{equation}
Z_L(p)=aM(p),\label{zl}
\end{equation}
which are positive and finite constants for $p=\pi_A$ (see eq.(\ref{m})). 
Together with the propagators (\ref{sn11}) obtained by the strong coupling 
expansion for $g_2\gg 1$ and $p\sim\pi_A$, we conclude that the doublers of
the neutral channel (\ref{rsn11}) are indeed relativistic massive particles, 
whose wave functions can be renormalized according to (\ref{rbound}).

Given the strong coupling $g_2\gg 1$, in spite of the propagator (\ref{rsn11}) 
for the neutral
three-fermion-state $\Psi_L^n$ resulting from the strong coupling 
expansion for $p\sim
\pi_A$, we can make an analytical continuation from $p\sim\pi_A$ to $p\sim 0$.
However, for the reason that $Z^2_L(p)\rightarrow O(p^8)$ as $p\rightarrow 0$, we cannot
conclude that $p\sim 0$ is a simple pole for a relativistic massless particle
by an analytical continuation of the propagator (\ref{rsn11})
from $p\sim\pi_A$ to $p\sim 0$, whereas we are not allowed to perform a
wave-function renormalization (\ref{rbound}) for $p\sim 0$. Actually, the
propagator (\ref{rsn11}) is vanishing at $p\sim 0$. It is important to point
out that at the limit of $p\rightarrow 0$, the vanishing of the propagator
(\ref{rsn11}) is definitely positive, i.e., it is a double zero. This
implies that ghost states with negative norm would not appear in low-energy
spectrum. 

However, on the other hand, we cannot conclude that the vanishing of the 
neutral propagator
(\ref{rsn11}) for $p\sim 0$ indicates the virtual state of three fermions in
the neutral channel. Since at $p=0$ the static limit is trivial, the computation of
the propagator (\ref{rsn11}) by the strong coupling expansion may not be reliable.
We must compute exactly the same two-point Green function
for the neutral three-fermion-operator, 
\begin{equation}
\int_x e^{-ipx}\langle\Psi^n_L(0)\bar\Psi^n_L(x)\rangle,
\label{3green}
\end{equation}
for $g_2\gg 1$ and $p\sim 0$, as that (\ref{rsn11}) computed by the strong coupling
expansion for $p\sim \pi_A$ and $g_2\gg 1$. Since the coupling $g_2$
in segment ${\cal A}$ can be arranged such that
\begin{equation}
a^2g_2w(p)< 1,\hskip0.7cm \infty >g_2\gg 1,\hskip0.4cm p\sim 0, 
\label{effweak}
\end{equation}
we can adopt the effective weak coupling expansion in 
powers of $a^2g_2w(p)$ to calculate the Green function (\ref{3green}).
This is due to the fact that $\chi_R$ is always an external field in 
Feyman diagrams (see Fig.3) of computing this Green function 
(\ref{3green}) and each 
$\chi_R$ is associated with $g_2w(p)$. This is a very 
reliable approximation as far as condition (\ref{effweak}) holds.
Here, we want to emphasize that the meaning of $p\sim 0$ is,
\begin{equation}
\epsilon > p\gg a\tilde v\simeq 0.
\label{p0}
\end{equation}
We can determine the value
of $g_2$ in such a way that the threshold scale $\epsilon$ is much larger than
the soft symmetry breaking scale $\tilde v$.

Clearly, no bound states in eq.(\ref{3green})
can be formed for such a weak effective coupling. Only the virtual state for
the neutral three-fermion-cut (\ref{cut}) can be found in this Green function 
(\ref{3green}). We show this by calculating (\ref{3green}) in
the effective weak coupling expansion as indicated in the Feyman diagrams 
(Fig.3). The leading order ($O(1)$) of this expansion is the first diagram in 
Fig.3, which corresponds to two parts,
\begin{eqnarray}
W_\circ (x) &=&-\left({1\over2a}\right)^2\langle\psi^{i\gamma}_L(0)\bar\psi^{j\delta}_L(x)\rangle
\langle\psi^\delta_R(0)\bar\psi^\gamma_R(x)\rangle
\langle\psi^{j\alpha}_L(0)\bar\psi^{i\beta}_L(x)\rangle,
\label{lcut}\\
W'_\circ (x) &=&\left({1\over2a}\right)^2\langle\psi^{i\gamma}_L(0)\bar\psi^\gamma_R(x)\rangle
\langle\psi^\delta_R(0)\bar\psi^{j\delta}_L(x)\rangle
\langle\psi^{j\alpha}_L(0)\bar\psi^{i\beta}_L(x)\rangle,
\label{lcut'}
\end{eqnarray}
where $\gamma,\delta,\beta,\alpha$ are spinner indices. In momentum space,
eqs.(\ref{lcut}) and (\ref{lcut'}) are given as,
\begin{eqnarray}
W_\circ(p) &=&-\int_{qk}S^{ji}_{LL}(p+q)\tr\left[S_{RR}(k-{q\over2})
S^{ij}_{LL}(k+{q\over2})\right]\left({1\over2a}\right)^2,
\label{cut1}\\
W'_\circ(p) &=&\int_{qk}S^{ji}_{LL}(p+q)\tr\left[\Sigma^i(k-{q\over2})
\right]\tr\left[\Sigma^j(k+{q\over2})\right]\left({1\over2a}\right)^2,
\label{cut1'}
\end{eqnarray}
where $p$ is the total external momentum, $k$ and $q$ are the relative 
internal momenta. Since 
segment ${\cal A}$ is an entirely symmetric phase, i.e.~$\Sigma^i(k)=0$ 
(see eqs.(\ref{ws2},\ref{ws2'})), eq.(\ref{cut1'}) identically vanishes. In general,
we can write the internal propagators 
$S^{ij}_{LL}(k)$ and $S_{RR}(k)$ in eq.(\ref{cut1}) as follow,
\begin{eqnarray}
S^{ij}_{LL}(k)&=&\delta_{ij}f_L(k^2)\gamma_\mu\sin k^\mu P_R,\label{lp}\\
S_{RR}(k)&=&f_R(k^2)\gamma_\mu\sin k^\mu P_L.
\label{rp}
\end{eqnarray}
These internal propagators $S_{LL}^{ij}(k)$ and $S_{RR}(k)$ with internal
momentum $k\in (0,\pi_A]$ can be given by eqs.(\ref{lp'},\ref{rp'}) 
calculated by the strong coupling expansion in segment ${\cal A}$.
We define,
\begin{equation}
R(k,q)= \tr\left[S_{RR}(k-{q\over2})
S^{ij}_{LL}(k+{q\over2})\right],
\label{rr}
\end{equation}
where we ignore indices $i,j$ in the LHS.
Since $R(k,q)$ (\ref{rr}) is an even function with respect to $k$ and 
$q$,
\begin{equation}
R(k,q)=R(-k,q),\hskip0.3cm R(k,q)=R(k,-q),
\label{r}
\end{equation}
one can easily show for the external momentum $p\sim 0$,
\begin{equation}
W_\circ(p) \simeq ac_\circ(i\gamma_\mu p^\mu) +O(p^2),
\label{w00}
\end{equation}
where $c_\circ$ is a constant.

The second diagram of Fig.3 denotes the contribution of order 
$O[(g_2w(p)]^2]$ to the Green function (\ref{3green}) given by\footnote{
The contribution (\ref{lcut'}) is zero, and other possible 
contributions are identically zero, owing to $\Sigma(k)=0$.}, 
\begin{equation}
W_1(p)=[g_2w(p)]^2V(p)\tilde S_{RR}(p)V(p),\label{w1}
\end{equation}
where $\tilde S_{RR}(p)$ is the full propagator of $\chi_R$, as indicated by a
full circle in the 
middle of the Feyman diagram (Fig.3), which is summed over all contributions of 
this effective weak coupling expansion (Fig.4). We can adopt eq.(\ref{rp'}) for 
$\tilde S_{RR}(p)$ by an analytical continuation from $p\sim\pi_A$ to $p\sim 0$.
On the other hand, as a consequence of the 
$\chi_R$-shift-symmetry (\ref{free}), $\tilde S_{RR}(p)$ for $p\sim 0$ is
a free propagator,
\begin{equation}
\tilde S_{RR}(p)\simeq {a\over i\gamma_\mu p^\mu}.
\label{free'}
\end{equation} 
In eq.(\ref{w1}), $V(p)$ is given by
\begin{equation}
V(p) =\int_{qk}4w(k-{q\over2})S^{ji}_{LL}(p+q)\tr\left[S_{RR}(k-{q\over2})
S^{ij}_{LL}(k+{q\over2})\right],
\label{v}
\end{equation}
where the factor $4w(k-{q\over2})$ comes from the interaction vertex.
Using a similar analysis to
that giving eq.(\ref{w00}), we get for $p\sim 0$ (up to a finite constant),
\begin{equation}
V(p)\sim -a(i\gamma_\mu p^\mu),
\label{v'}
\end{equation}
which cancels the pole of eq.(\ref{free'}) at $p\sim 0$. As a result,
we obtain ($w(p)\sim p^2$, $p\sim 0$),
\begin{equation}
W_1(p) \simeq (a^2g_2p^2)^2ac_1(i\gamma_\mu p^\mu),
\label{w11}
\end{equation}
where $c_1$ is a finite number. Thus, for $g_2\gg 1$ and $p\sim 0$, the Green
function (\ref{3green}) is,
\begin{equation}
\int_x e^{-ipx}\langle\Psi^n_L(0)\bar\Psi^n_L(x)\rangle\simeq W_\circ(p) + W_1(p),
\label{3cut}
\end{equation}
which is regular at $p\sim 0$. This shows that the 
neutral 
channel is a virtual state for the three-fermion-cut in region (\ref{effweak}).
Obviously, it is absolutely 
incorrect for doublers ($p\sim\pi_A$) since the effective coupling 
(\ref{effweak}) can be extremely
large, bound states $\Psi_L^n(x)$ (\ref{bound}) must be formed.  

Note that in this region (\ref{effweak}), the mixing between the elementary 
field $\chi_R$ and 
neutral three-fermion-state $\Psi^{n}_L$ (\ref{bound}) calculated
by the strong coupling expansion for $p\sim \pi_A$ is given by (\ref{mx}).
Analytical continuation of (\ref{mx}) from $p\sim\pi_A$ to $p\sim 0$ shows
the vanishing of the mixing at $p\sim 0$, while this mixing
gives rise to the gauge-invariant mass terms for doublers $p\sim \pi_A$.
On the other hand, the mixing (\ref{mx}) can be calculated by
the effective weak coupling expansion (\ref{effweak}). The leading
order is,
\begin{equation}
\int_xe^{-ipx}\langle \Psi^{n}_L(0)\bar\psi_R(x)\rangle =\int_xe^{-ipx}
\left({1\over a}\right)\langle\psi_L^i(0)\bar\psi_R(x)\rangle
\langle\psi_R(0)\bar\psi^i_L(0)\rangle + O(g_2w(p)),
\label{cut3}
\end{equation}
which vanishes for no hard spontaneous symmetry breaking.

Based on the computations of the Green function for the neutral channel 
(\ref{3green}) at both $p\sim\pi_A$ (\ref{rsn11}) and $p\sim 0$ (\ref{3cut}) for $g_2\gg 1$, we 
conclude that in the intermediate region (\ref{effweak}) of the gauge-symmetric
segment ${\cal A}$, the low-energy spectrum ($p\sim 0$) is 
\begin{itemize}
\item
undoubled for all doublers ($p\sim \pi_A$) decoupled; 
\item 
chiral because for only exists a free 
right-handed mode $\chi_R$, while $\Psi^n_L$ is no longer a bound
state, it is instead a virtual state ${\cal C}[\Psi_L^n]$.

\end{itemize}
By the analytical continuity of the Green function 
(propagator) (\ref{3green}) in terms of
the total momentum $p$, from eq.(\ref{rsn11}) for $p\sim \pi_A$ to
eq.(\ref{3cut}), we must find the threshold scale $\epsilon$ (\ref{disscale}),
where the neutral three-fermion-state $\Psi^n_L(x)$ (\ref{bound}) with the ``wrong'' chirality
dissolves into its virtual states and only $\chi_R$ remains as a relativistic
particle at $p\sim 0$. We emphasize that this already violates the ``no-go'' 
theorem even though the chiral fermion is neutral. However, it is worthwhile
to point out that such a mechanism violating the ``no-go'' theorem is only 
expected to work in the cases of neutral and anomaly-free theories.

\section{The three-fermion-cut for the charged channel}

The form factor $Z_R(p)$ (\ref{zlzr}) for the right-handed
three-fermion-state $\Psi_R^i(x)$ (\ref{bound}) can not be determined by the
Ward identity (\ref{w}). Instead, it can be calculated by using the results
obtained in the strong coupling expansion for $g_2\gg 1$ and $p\sim \pi_A$. 

The 1PI-vertex function associated to $Z_R(p)$ is given by the truncated
Green function (\ref{zlzr}) that is defined as (as indicated in Fig.2),
\begin{eqnarray}
Z_R(p)&=&\int_xe^{-ipx}{\delta^{(2)}\Gamma\over\delta\Psi'^i_R(0)
\delta\bar\psi'^j_L(x)}\nonumber\\
&=&\int_xe^{-ipx}\int_{z_1,z_2}
\left(G^{jl}_{MM}(x,z_2)\right)^{-1}
G^{lk}_{ML}(z_2,z_1)
\left(G^{ki}_{\rm free}(z_1,0)\right)^{-1}\nonumber\\
&=&\left(S^{jl}_{MM}(p)\right)^{-1}
S^{lk}_{ML}(p)
\left(S^{ki}_{\rm free}(p)\right)^{-1},\label{truncated}
\end{eqnarray}
where two-point Green functions are given by,
\begin{eqnarray}
G^{jl}_{MM}(x,z_2)&=&\langle\Psi_R^j(x)\bar\Psi_R^l(z_2)\rangle\rightarrow
S^{jl}_{MM}(p),\nonumber\\
G^{lk}_{ML}(z_2,z_1)&=&\langle\Psi_R^l(z_2)\bar\psi_L^k(z_1)\rangle
\rightarrow S^{lk}_{ML}(p'),\nonumber\\
G^{ki}_{\rm free}(z_1,0)&=&\langle\psi_L^k(z_1)\bar\psi_L^i(0)\rangle_\circ
\rightarrow S^{ki}_{\rm free}(p''),
\label{greens}
\end{eqnarray}
and their transformations into the momentum space in the last line of 
eq.(\ref{truncated}), where $S^{ki}_{\rm free}(p'')$ is the free propagator of of the 
$\psi^i_L$ in the absence of multifermion couplings $g_1,g_2$.

Adopting the results $S^{jl}_{MM}(p)$ (\ref{sc11}), $S^{ij}_{ML}(p)$
(\ref{mixingc}), which is obtained by the strong coupling expansion, and the inverse free propagator $(S^{ki}_{\rm free}(p))^{-1}=\left({i\over a}\right)\sum_\mu p^\mu\gamma_\mu$ of the 
$\psi^i_L$, we get,
\begin{equation}
Z_R(p)=aM(p),\label{zr}
\end{equation}
which is the same as $Z_L(p)$ (\ref{zl}) directly derived from the Ward identity (\ref{w}). 
Eq.(\ref{zr}) is a positive and finite constant for doublers $p\simeq\pi_A$ 
(see eq.(\ref{m})). 
Together with the propagator (\ref{rsc11}) obtained by the strong coupling 
expansion for $g_2\gg 1$ and $p\sim\pi_A$, we conclude that the doublers of
charged channel (\ref{rsc11}) are indeed relativistic massive particles, 
whose wave functions can be renormalized according to (\ref{rbound}).

In spite of  eqs.(\ref{rsc11}) and (\ref{zr}) obtained by the strong coupling 
($g_2\gg 1$) expansion for $ p\sim\pi_A$, we can analytically continue the 
momentum ``$p$'' in these 
equations to the limit of $p\rightarrow 0$. When $p\rightarrow 0$,
$Z^2_R(p)\rightarrow O(p^8)$, the propagator (\ref{rsc11}) of charged
three-fermion-state vanishes. This implies that the low-energy state 
($p\sim0$) of $\Psi^i_R(x)$ with the ``wrong'' chirality
is no longer a simple pole as a relativistic particle, instead, it is a virtual
state,\footnote{ 
We emphasize again that at the limit of $p\rightarrow 0$, the vanishing of
the propagator of the charged three-fermion-state is definitely positive, 
i.e., it is a double zero, which implies
that ghost states with negative norm do not appear and couple to the gauge field
in the low-energy limit. Otherwise, the theory would be inconsistent
\cite{incon}.} and the low-energy state of chiral fermions with ``right'' chirality
should be realized in the spectrum of charged particles. We have to confess that unlike 
neutral channel case, without
further more rigorous computations, numerical one for instance, we can not demonstrate 
such a scenario for realizing charged lattice chiral fermions to be definite ture, since 
the static limit is broken at $p=0$. However, it is strongly emphasized that for the same 
reason that the static limit is broken at $p=0$, we can not conclude the spectrum to be 
definitely vector-like as it was
concluded in ref.\cite{sy93}. What we can conclude is a loophole at $p\sim 0$ opening up.
We go further in following to check in more details.

In order to directly show
that the low-energy state ($p\sim 0$) of $\Psi^i_R(x)$ is the virtual state for a
three-fermion-cut. We have to compute exactly the same two-point Green function, 
\begin{equation}
\int_x e^{-ipx}\langle\Psi^i_R(0)\bar\Psi^j_R(x)\rangle,
\label{3greenc}
\end{equation}
of the charged channel for $p\sim 0$ and $g_2\gg 1$ as that (\ref{rsc11})
computed by the strong coupling expansion for $p\sim\pi_A$ and $g_2\gg 1$.
However, unlike the case of neutral channel, we do not have the reliable 
method of the effective weak coupling expansion in 
powers of $a^2g_2w(p)$ (\ref{effweak}), since $\psi^i_L(x)$ is always an external 
field in the Feyman diagrams computing the Green function (\ref{3greenc}) and each 
$\psi_L$ does not 
associate with $g_2w(p)$. In this case, the effective weak coupling expansion 
used for computing the neutral channel (Fig.3) must breakdown for $g_2\gg 1$.

Nevertheless, we observe that the last binding threshold is located at 
$a^2g_c^a=0.124$ (\ref{segment}). Above this threshold 
($g_2>g_c^a, a^2g_2\sim O(1)$ eq.(\ref{o(1)})),
all doublers are supposed to be decoupled via eqs.(\ref{sc11},\ref{sn11}) and 
(\ref{m},\ref{di}).
The critical point $a^2g^a_c=0.124$ is a rather small number. 
Thus, at $a^2g_2\sim O(1)$, we introduce $N_f$, an addition number of fermion 
flavours of $\psi^i_L$ and $\chi_R$ so that,
\begin{equation}
a^2g_2>0.124,\hskip0.5cm \tilde g_2= a^2g_2N_f<1\hskip0.5cm {\rm fixed },
\hskip0.5cm 
N_f=3\sim 8,
\label{largen}
\end{equation}
where the value of $N_f$ depends on the value of $a^2g_2$ considered.
Therefore, in a certain intermediate region of $a^2g_2\sim O(1)$, we adopt
the large-$N_f$ technique to control the convergence of the approximation (see
Fig.3) and calculate the Green function (\ref{3greenc}) of the charged
three-fermion-operators for the low-energy $p\sim 0$.
Hence we can get a qualitative insight into the charged low-energy spectrum ($p\sim
0$) within the intermediate region of $g_2$ (\ref{largen}). We expect that the dynamics of
the interaction is not greatly changed by introducing more flavours. As for 
$a^2g_2> 1$, such large-$N_f$ technique is doomed to fail.

Analogous to the case of neutral channel (\ref{w00}), the non-trivial
leading order $O(N_f)$ contribution, as indicated by the first diagram in 
Fig.3, is ($p\sim 0$),
\begin{equation}
W^c_\circ(p) = -N_f\int_{qk}S_{RR}(p+q)R(k,q)\left({1\over2a}\right)^2.
\label{cut1c}
\end{equation}
The second order $(a^2g_2N_f)^2$ contribution, as indicated by the second 
diagram in Fig.3, is ($p\sim 0$),
\begin{equation}
W^c_1(p)=(g_2N_f)^2V^c(p)\tilde S_{LL}(p)V^c(p)\label{wc1}
\end{equation}
where we ignore indices $ij$ and $V^c(p)$ is given as ($p\sim 0$)
\begin{equation}
V^c(p) =-\int_{qk}4w(p+q)w(k-{q\over2})S_{RR}(p+q)R(k,q),
\label{vc}
\end{equation}
where the factor $4w(p+q)w(k+{q\over2})$ comes from interacting vertices.
The full propagator $\tilde S_{LL}(p)$ of 
$\psi_L^i(x)$ in eq.(\ref{wc1}), as indicated by a full circle in middle of the 
second diagram in Fig.3, is calculated by  the train approximation 
(as indicated in Fig.4) for the external total momentum $p\sim 0$,
\begin{equation}
\tilde S^{ij}_{LL}(p)=\int_xe^{-ipx}\langle\psi_L^i(0)\bar\psi^j_L(x)\rangle\simeq
 Z_2^{-1}(p)S^{ij}_{LL}(p).
\label{lpro}
\end{equation}
The wave-function renormalization $Z_2(p)$ of the elementary field 
$\psi_L^i(x)$ is given by (see Appendix A),
\begin{equation}
Z_2(p)\!=\!1\!+\!{(\tilde g_2)^2\over N_f}\!\int_{k,q}\!\left(\!4
w(p+q)w(k\!-\!{q\over2})\right)^2\!S_{RR}(p+q)S_{LL}(p)R(k,\!q).
\label{z2p}
\end{equation}
Note that in eqs.(\ref{cut1c},\ref{vc},\ref{z2p}) and $R(k,q)$ (\ref{rr}), the
internal propagators $S_{LL}(k)$ for $\psi_L^i(x)$, and $S_{RR}(k)$ for
$\chi_R(x)$ 
are respectively given by
eqs.(\ref{lp},\ref{rp}) or approximately by eqs.(\ref{lp'},\ref{rp'}). The
reasons for such choices are that in the region (\ref{largen}), doublers
($k\sim \pi_A$) are supposed to be decoupled via eq.(\ref{lp'},\ref{rp'}) and
the internal momentum $k$ runs from $k\sim 0$ to $k\sim \pi_A$. As for the
propagator $S_{LL}(p)$ in eqs.(\ref{lpro},\ref{z2p}) for the
external total momentum $p\sim 0$, we adopt eq.(\ref{lp'}) by analytical 
continuation from $p\sim \pi_A$ to $p\sim 0$.

Analogous to that (\ref{w00}) in the neutral channel,
for the total momentum $p\sim 0$, the leading order contribution (\ref{cut1c})
becomes\footnote{``$\sim$'' indicates up to a finite constant.},
\begin{equation}
W^c_\circ(p) \sim aN_f(i\gamma_\mu p^\mu).\label{w00'}
\end{equation}
As for the second order contribution (\ref{wc1}), we find for $p\sim 0$,
\begin{equation}
V^c(p)\sim a(i\gamma_\mu p^\mu),
\label{vc'}
\end{equation}
which cancels the pole at $p=0$ stemming from the propagator $\tilde S_{LL}$
in eq.(\ref{lpro}). And the wave-function renormalization $Z(p)$ of the 
elementary fields $\psi_L^i(x)$ at $p=0$ is given by,
\begin{equation}
Z_2(0)= 1+ {\rm const.},
\label{z2}
\end{equation}
due to eq.(\ref{rr}). This indicates that the relativistic particle ($p=0$) of 
$\psi^i_L$ receives a wave-function renormalization $Z_2(0)$.
As a result, the second-order contribution (\ref{wc1}) reads,
\begin{equation}
W^c_1(p) \sim (\tilde g_2)^2a(i\gamma_\mu p^\mu).
\label{w11'}
\end{equation}
Thus, for $p\sim 0$ the Green function (\ref{3greenc}) is approximately
computed as,
\begin{equation}
\int_x e^{-ipx}\langle\Psi^c_R(0)\bar\Psi^c_R(x)\rangle\simeq W^c_\circ(p) 
+ W^c_1(p),
\label{3cut'}
\end{equation}
which is regular at $p\sim 0$. This implies that in this intermediate region
(\ref{largen}), the charged channel of three-fermion-operators at $p\sim 0$ is
not a simple pole for a massless relativistic particle with the ``wrong'' chirality,
rather it is regular for a virtual state of three-fermion-cut. This agrees with
the result obtained by analytical continuation of the propagator (\ref{rsc11}) 
from $p\sim\pi_A$ to $p\sim 0$.

An important consistent check is to examine the equation (\ref{wc1}) for the
``doublers'', i.e., the external momentum $p\sim\pi_A$. In fact, for
$p\sim\pi_A$, 
\begin{equation}
V^c(p)\simeq -4w(\pi_A)\int_{qk}w(k-{q\over2})S_{RR}(p+q)R(k,q).
\label{vcp}
\end{equation}
This results in the coupling $\tilde g_2$ in the second-order contribution
(\ref{wc1}) being enhanced up a factor of $w^2(\pi_A)$ (see eq.(\ref{wisf})),
\begin{equation}
W^c_1(p)\sim (\tilde g_2w(\pi_A))^2V^c(p)\tilde S_{LL}(p)V^c(p).
\label{wc1p}
\end{equation}
Analogously, for $p\sim\pi_A$,
\begin{equation}
Z_2(p)\simeq 1+{2(\tilde g_2w(\pi_A))^2\over N_f}\int_{k,q}\left(4
w(k-{q\over2})\right)^2S_{RR}(p+q)S_{LL}(p)R(k,q).
\label{z2pp}
\end{equation}
The consequence is the complete breakdown of the large-$N_f$ expansion 
(\ref{largen}), which is not convergent for $p\sim\pi_A$, to calculate 
the two-point Green function (\ref{3greenc}) of charged three-fermion-operators.
This implies that bound states (three-fermion-states) with the total momentum
$p\sim\pi_A$ should be formed,
consistently with the bound states that we find by the strong coupling 
expansion for $p\sim \pi_A$. 

We turn to the computation of Green function for the mixing between the 
elementary field $\psi^i_L$ and charged three-fermion-state $\Psi^j_R$ 
(\ref{bound}). This mixing can be calculated by
the large-$N_f$ expansion (\ref{largen}). The non-trivial leading
order $O(N_f)$ is explicitly written as,
\begin{equation}
\int_xe^{-ipx}\langle \Psi^i_R(0)\bar\psi^j_L(x)\rangle =
\int_xe^{-ipx}\left({1\over a}\right)\langle\psi_R(0)\bar\psi^j_L(x)\rangle
\langle\psi^i_L(0)\bar\psi_R(0)\rangle +O\left({\tilde g_2\over N_f}\right),
\label{cut3'}
\end{equation}
where $O\left({\tilde g_2\over N_f}\right)$ stands for higher order 
contributions.
Eq.(\ref{cut3'}) vanishes for non spontaneous symmetry breaking 
(see eq.(\ref{ws2})). Consistently, an analytical continuation of the Green 
function (\ref{mixingc}) 
from $p\sim\pi_A$ to $p\sim 0$ shows vanishing of the mixing at $p\sim 0$
as well.

Finally we check that in the intermediate region of $g_2$ (\ref{largen}),
whether the Green function (\ref{lpro}) for the elementary field $\psi_L^i$ has
a simple pole at $p\sim 0$ representing a relativistic massless particle. Due
to $Z_2(0)=1+ {\rm const.}$ (\ref{z2}), one can check the propagator
(\ref{lpro}) for $\psi_L^i$ has a simple pole at $p=0$, which indicates a
massless, charged left-handed $\psi_L^i$ in the low-energy spectrum. This
agrees with the analytical continuation of the propagator (\ref{lp'}) from
$p\sim\pi_A$ to $p\sim 0$. 

In the continuation from eq.(\ref{rsc11}) for $p\sim \pi_A$ to eq.(\ref{3cut'})
for $p\sim 0$, we must meet the ``dissolving'' threshold $\epsilon$, which
should be the same as that in the neutral channel. We need to point out that
the computation of charged channel is different from the computation of neutral
channel. In the neutral channel, the Green functions 
(\ref{rp'},\ref{mx},\ref{rsn11})) for
$p\sim\pi_A$ and (\ref{free'},\ref{3cut},\ref{cut3}) for $p\sim 0$ are both
consistently calculated in $g_2\gg 1$. 
However, in the charged channel, the Green functions (\ref{lpro},\ref{3cut'},
\ref{cut3'})
for $p\sim 0$ are computed in the intermediate region $a^2g_2\sim O(1)$
(\ref{largen}), while
the same Green functions (\ref{lp'},\ref{rsc11},\ref{mixingc}) for $p\sim\pi_A$ are 
computed in the region
$g_2\gg 1$. This may raise a question whether the dynamics we explore for
$p\sim 0, a^2g_2\sim O(1)$ and for $p\sim\pi_A, a^2g_2\gg 1$ are consistent.
We argue that such studies are qualitatively justified, since the computations
in the strong coupling expansion with respect to doublers ($p\sim\pi_A$) are
valid as well for $a^2g_2\sim O(1)$ (\ref{o(1)}) as discussed in the section II.

In conclusion, on the basis of approximate calculations of relevant two-point
Green functions of elementary and composite three-fermion-operators with respect to
doublers $p\sim\pi_A$ and low-energy mode $p\sim 0$, we qualitatively explore 
a possible scaling window in segment ${\cal A}$, where
in the low-energy spectrum, the three-fermion-states with the ``wrong''
chirality turn into their corresponding virtual states, and elementary fermion
states with the ``right'' chirality remain as massless states. 
Evidently, full non-perturbative numerical simulation to explore
this scaling window is very inviting and necessary in particular
for any solid conclusions in the charged channel. 

\section{Discussions of relations between our model, the Eichten-Preskill 
and Smit-Swift models}

Multifermion coupling models, for instance the Eichten-Preskill model and 
the one presented in this paper,
can formally be rewritten as scalar-fermion coupling models in the lattice (cutoff) 
scale by introducing 
appropriate scalar fields, which mimic the bound states of two fermions.  
There are many classes of
multifermion coupling models and scalar-fermion couplings models, the multifermion 
coupling models and scalar-
fermion coupling models belong to the same universal class and are certainly equivalent in the sense that not 
only they can be formally translated each other by introducing auxiliary scalar fields in path integrals, but
also they share the same infrared fix (scaling) points and lines in the phase space and have the same low-energy 
spectra, couplings and relevant operators upon those fix points and lines. The later property is more important
and difficult to generally prove the equivalence between a given multifermion model and a given
scalar-fermion coupling model. What we can do for a given either multifermion coupling model or scalar-fermion coupling model is to analyze all possible infrared scaling regin in phase space. 

The Smit-Swift model certainly belongs to one universal class of scalar-fermion coupling models and equivalent
multifermion coupling models. The Eichten-Preskill multifermion-coupling model and its equivalent scalar-fermion
coupling model could belong to another universal class that is not proved to be the same as the universal
class of models of the Smit-Swift type. It is no doubt that our 
multifermion- coupling model has its equivalent scalar-fermion coupling model. However, no one has demonstrated that our multifermion-coupling model must be in the same universal classes of either the Smit-Swift model or the Eichten-Preskill model. The only thing
to do is to analyze all possible scaling region of our model by appropriate methods as that we present in this paper. We can clearly point out why and what could make possible differences between our model and the Eichten-Preskill as well as Smit-Swift models in the existence of a possible scaling region for lattice chiral fermions.

The common feature of our model, the Eichten-Preskill model and Smit-Swift model is
the momentum-dependence of multifermion couplings or scalar-fermion couplings attempting to
give doublers $p=\pi_A$ heavy masses and decouple them from the low-energy spectrum. In the phases of weak multifermion couplings or scalar-fermion couplings, doublers are not decoupled and the spontaneous symmetry breaking could occur. Therefore, in these three models we only need to discuss the phase of strong multifermion couplings or scalar-fermion couplings where three-fermion-states or bound states of scalar-fermion are formed, and spectrum is expected to be vector-like. Question clearly is whether we can find a scaling region in the strong-coupling phase of these three models, and in this scaling region the low-energy mode of three-fermion-states 
or scalar-fermion bound states with the ''wrong'' chirality can be excluded, and the chiral fermions with the ''right'' chirality exsit in the low-energy spectrum.

In the strong-coupling phase, the momentum-dependence of the strong multifermion coupling of our model is elaborately made in such a way that in a segment of the strong coupling phase (${\cal A} g_1=0, g_2\gg 1$) for all doublers $p=\pi_A$ the 
strong-coupling limit (static limit) exists and for the low-energy mode $p\sim 0$ the strong-coupling limit does not exist as discussed in section III, beside no hard spontaneous symmetry breaking occurs as discussed in eqs.(\ref{ws2},\ref{ws2'}) and ref.\cite{xue97}. These crucial features are very different from the momentum-dependence of the Eichten-Preskill and Smit-Swift models in their strong-coupling phases, where no such segment have been found. As the consequences of these features, our model has the following properties differentiated from the Eichten-Preskill and Smit-Swift models.   

In the strong coupling phase (PMS) of the Smit-Swift model, only neutral sector of vector-like fermion exits in the spectrum which is computed by the strong-coupling expansion\cite{gw} method analogous to the one adopted in this paper. The
wave-function renormalization of the composite neutral field is given by the
vacuum expectation value of scalar fields $V(x)$, $z^2=\langle
V^\dagger(x)V(x+a)\rangle$\cite{gw}, which does not vanish as $p\rightarrow 0$, showing 
the neutral scalar-fermion bound state can not be excluded from the low-energy spectrum. However, as clearly demonstrated in section V, the neutral three-fermion-state, which dissolves into the three-fermion-cut, is excluded from the low-energy spectrum for its wave-function renormalization goes to zero as $p\sim 0$ and effective multifermion coupling $g_2(p)$ vanishes 
without any hard spontaneous symmetry breaking. Although the 
scalar-fermion coupling in the Smit-Swift model can be reduced so that the scalar-fermion bound state could disappear, the model runs into the phase of hard spontaneous symmetry breaking ($v\sim O({1\over a})$). 
    
In the strong coupling phase of the Eichten-Preskill model, not only neutral sector but also charged sector of vector-like fermions exits in the spectra, both are computed by the strong-coupling expansion\cite{ep,gpr} method that is the exactly same as adopted in this paper. However, the strong coupling limit (the static limit) of the Eichten-Preskill model is regular at $p\sim 0$ differentiated from the singularity of the static limit at $p\sim 0$ of our model.
This implies that the multifermion couplings as function of the momentum $p$ of the Eichten-Preskill model are not appropriate so as to find a scaling region,
these massive charged fermions remain as vector-like not only for $p\sim\pi_A$ modes and but also for 
the low-energy mode ($p\sim 0$). Beside, the multifermion couplings as function of the momentum $p$ of the Eichten-Preskill model develop a hard spontaneous symmetry breaking in an expected scaling region\cite{gpr}. The multifermion couplings as function of the momentum $p$ of our model are carefully made to avoid the hard spontaneous symmetry breaking in the possible scaling region ${\cal A}$, the shift-symmetry (\ref{shift}) plays extremely important role in this matter. The Eichten-Preskill model for $SU(5)$ and $SO(10)$ would have a chance to work, if its multifermion couplings are more carefully defined.

\section{Some remarks}

In this paper, we have discussed the features of the spectrum of neutral
and charged sectors appearing in segment ${\cal A}$ (\ref{segment}). It is interesting to
point out that in this scenario, doublers ($p\sim\pi_A$) are decoupled by a
gauge-invariant mass term, while the low-energy modes with the ``wrong'' chirality 
are ``decoupled'' by the vanishing of their generalized form factor. However, 
it is still far from a definitive demonstration 
that chiral gauge theories in the low-energy limit can be achieved in this way,
we need to have numerical
simulations to show that this scenario is indeed realized in segment 
${\cal A}$. 

The whole spectrum in segment ${\cal A}$ is gauge symmetric, and Ward 
identities of gauge symmetry
are preserved. We can straightforwardly turn on the perturbative gauge interaction.
By the strong coupling expansion in powers of ${1\over g_2}$, we 
compute\cite{xue97l}
three-point vertex function $\langle\Psi^i_R(x)\bar\Psi^j_R(y) A_\mu(z)
\rangle$ and obtain the vertex of the $SU_L(2)$ gauge field coupling to
the charged three-fermion-state (the leading order of gauge coupling $O(g)$),
\begin{equation}
\Lambda^{(1)}_{\mu RR}(p,p') =ig{\tau^a\over2}
\gamma_\mu P_R \cos{(p+p')\over2},
\label{gauge}
\end{equation}
where the momenta of three-fermion-states $p,p'\sim\pi_A$.
According the renormalization (\ref{ren}) of truncated Green functions with two
three-fermion-operator insertions, 
the vertex of gauge coupling to the three-fermion-states is given by (Fig.5),
\begin{equation}
\Lambda^{(1)}_{\mu RR}(p,p') =ig{\tau^a\over2}
\gamma_\mu P_R \cos{(p+p')\over2}Z_R(p)Z_R(p').
\label{rgauge}
\end{equation}
For $p\sim\pi_A$ and $p'\sim\pi_A$, $Z_R$'s are positive definite constants, we thus
renormalize the wave functions with respect to each doubler (\ref{rbound}), as discussed in 
section IV. As a result, 
gauge coupling vertex (\ref{rgauge}) turns into (\ref{gauge}). 
Although eq.(\ref{rgauge}) is obtained for $p,p'\sim\pi_A$ and $a^2g_2\gg 1$, it
can be analytically continued to $p,p'\sim 0$.
We find that in the limit of
$p\rightarrow 0$ and $p'\rightarrow 0$, the coupling vertex (\ref{rgauge}) of
three-fermion-operators and gauge boson vanishes as $O(p^8)$. This
consistently corresponds to the dissolving of three-fermion-states into 
three-fermion-cuts in the low-energy limit. Since the propagator of charged three-fermion-states 
{\it positively} vanishes, there are no ghost states with negative norm coupling to 
gauge field through the Ward identity stemming from the gauge symmetry in 
segment ${\cal A}$. 

The model presented in this paper cannot be considered as a realistic 
model reflecting all aspects of chiral gauge theories. We need to completely 
understand the relationship between the anomaly-free condition
and the realization of such dynamics discussed in the paper. We also need to 
understand what kind role of 't Hooft condition for
anomaly matching\cite{tha}, fermion-number violation and 
Witten's $SU(2)$ global anomaly\cite{witten} would play in such dynamics.
 
In this approach, the left-handed field $\psi^i_L(x)$ is the doublet of the
$SU_L(2)$ chiral gauge group. However, it can be generalized to be the left-handed
field (complex representations) of any anomaly-free chiral gauge group
(e.g.~$SU(5)$ and $SO(10)$).
A right-handed field (spectator) $\chi_R$, that is a singlet of the
chiral gauge group, can be introduced and coupled to the left-handed field in
the same way as the multifermion couplings given in (\ref{action}). As for the
right-handed field $\psi^i_R$ of chiral gauge groups, we can analogously
introduce a left-handed spectator field $\chi_L$ that is a singlet of the chiral
gauge groups, and couple it to the right-handed field $\psi^i_R$ in the same way as the multifermion
couplings in the action (\ref{action}) with $L\leftrightarrow R$. This indicates
that such a formulation of chiral gauge theories is actually quite general.

To be more
specific, we take the anomaly-free chiral gauge group of the Standard Model as
an example. In this realistic case, $\psi_L^i(x)$ can be both left-handed
lepton doublets and left-handed quark doublets. The candidate for a right-handed
spectator field $\chi_R$ could be the right-handed neutrino $\nu_R$. As for the
right-handed fields $\psi^i_R$ with respect to $U_Y(1)$, we can introduce an
additional left-handed spectator field $\chi_L$ (a $SU_L(2)\otimes U_Y(1)$
singlet) coupling to the right-handed fields $\psi_R^i$ as that in 
eq.(\ref{action}). These
spectator fields $\nu_R$ and $\chi_L$ are free and decouple from other
particles due to the $\nu_R$- and $\chi_L$-shift-symmetry acting on them. In
this way, we can in principle have a gauge-invariant formulation of the
Standard Model on the lattice. In practice, non-perturbative analysis, which
can be done analytically to show whether such a formulation gives the low-energy
Standard Model, should be more or less similar to (certainly more complicated than) that
discussed in this paper and references\cite{xue97,xue97l}. 
However, the spectator field $\chi_L$ might not be
necessary. Alternatively, the 't Hooft vertex in the lattice formulation of the
Standard Model\cite{mc} provides a scenario in which fermion-number
conservation is explicitly violated, and all chiral fermions
find their patterns with opposite chirality within the Standard Model. 
The formulation of a realistic Standard Model with
multifermion couplings and all features of dynamics discussed in this
paper are an extremely interesting subject for future studies. 

I thank Profs. G.~Preparata, M.~Cruetz, A.~Slavnov, E.~Eichten, and 
Jean Zinn-Justin for many helpful discussions.

\vskip1.5cm
\noindent
\appendix {\bf Appendix A}
\vskip1cm
The propagator of $\psi_L^i$ is given by eq.(\ref{lp'}) ($p\sim 0$),
\begin{equation}
S_{LL}^{ij}(p)=\delta_{ij}P_L\hat pP_R,\hskip1cm \hat p={{i\over a}\sum_\mu\gamma^\mu
\sin p_\mu\over {1\over a^2}\sum_\mu\sin^2p_\mu + M^2(p)}.
\nonumber
\end{equation}
The Feyman diagram (see Fig.~6) is given by,
\begin{eqnarray}
\hat\sigma^{ij}(p)&=&\delta^{ij}P_R\sigma(p)P_L\nonumber\\
\sigma(p)&=&-N_f\int_{qk} \lambda S_{RR}(p+q) \tr\left[S_{RR}(k-{q\over2})
S_{LL}(k+{q\over2})\right]\nonumber\\
&&=-N_f\int_{k,q}\lambda S_{RR}(p+q) R(k,q),
\label{sigma}
\end{eqnarray}
where $S_{LL}, S_{RR}$ are (\ref{lp'},\ref{rp'}) and $R(k,q)$ is 
eq.~(\ref{rr}) and 
\begin{equation}
\lambda=\left(4g_2w(p-k)w(k+{q\over2})\right)^2
={1\over N_f^2}a^{-4}\left(4\tilde g_2w(p-k)w(k+{q\over2})\right)^2.
\nonumber
\end{equation}
The wave-function renormalization $Z_2(p)$ of $\psi^i_L(x)$ in
eq.~(\ref{lpro}) can be calculated by using the train approximation 
(see Fig.~4),
\begin{eqnarray}
Z_2^{-1}S_{LL}^{ij}(p)&=&P_L(\hat p+\hat p\sigma\hat p+\hat p\sigma\hat p
\sigma\hat p+\cdot\cdot\cdot)P_R\delta^{ij}
\nonumber\\
&=&S_{LL}^{ij}(p)\left({1\over 1-\sigma\hat p}\right),
\label{train}
\end{eqnarray}
and one gets 
\begin{equation}
Z_2=1-\sigma\hat p.
\label{az2}
\end{equation}
With eq.~(\ref{sigma}) one can get 
\begin{equation}
\sigma\hat p=-
{1\over N_f}\!\int_{k,q}\!\left(4
\tilde g_2w(p-k)w(k\!+\!{q\over2})\right)^2S_{RR}(p+q)S_{LL}(p)R(k,q).
\label{az1}
\end{equation}
By substituting eq.~(\ref{az1}) into (\ref{az2}), one gets eq.~(\ref{z2p}).

\section*{Figure Captions} 
 
\noindent {\bf Figure 1}: \hspace*{0.2cm} 
The two-point Green functions of composite three-fermion-operators at 
strong coupling $g_2\gg 1$ for $p\sim \pi_A$, indicating three-fermion-states.

\noindent {\bf Figure 2}: \hspace*{0.2cm} 
1PI truncated Green functions of an elementary field and one 
three-fermion-operator insertion, i.e., the generalized form factors of 
three-fermion-states.

\noindent {\bf Figure 3}: \hspace*{0.2cm} 
The two-point Green functions of composite three-fermion-operators in the 
effective weak coupling for $p\sim 0$, indicating three-fermion-cuts. 

\noindent {\bf Figure 4}: \hspace*{0.2cm} 
The train approximation to the propagators of 
the elementary fields $\psi_L^i$ and $\chi_R$.

\noindent {\bf Figure 5}: \hspace*{0.2cm} 
Gauge boson coupling to three-fermion-states.

\noindent {\bf Figure 6}: \hspace*{0.2cm} 
A single bubble diagram in the weak-coupling expansion (see Figs.3,4).

\newpage 

\begin{figure}[t]
\centerline{
\psfig{figure=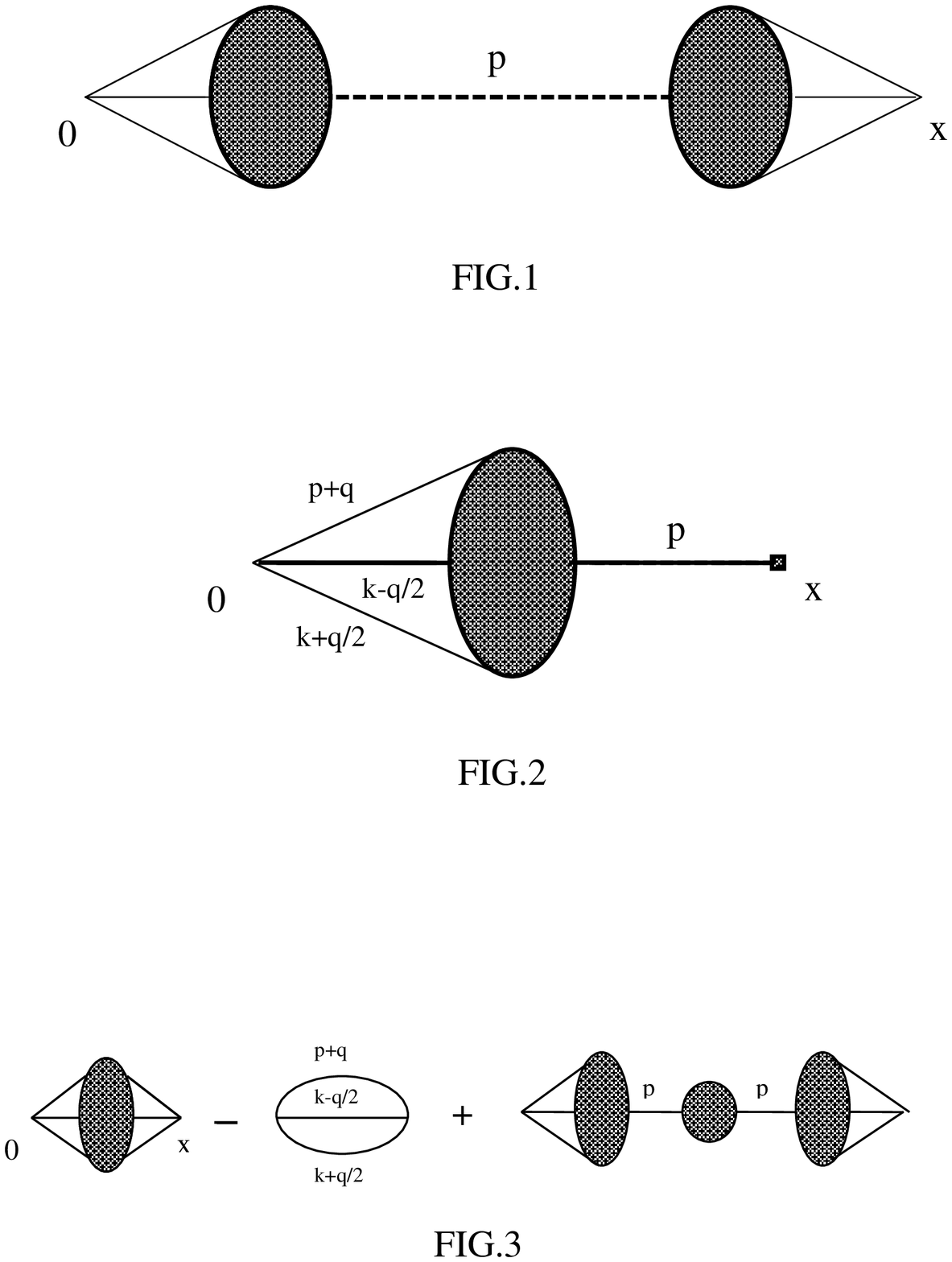}
}
\caption[]{Fig.1, Fig.2 and Fig.3}
\label{fig: fig3}
\end{figure}

\newpage 
\begin{figure}[t]
\centerline{
\psfig{figure=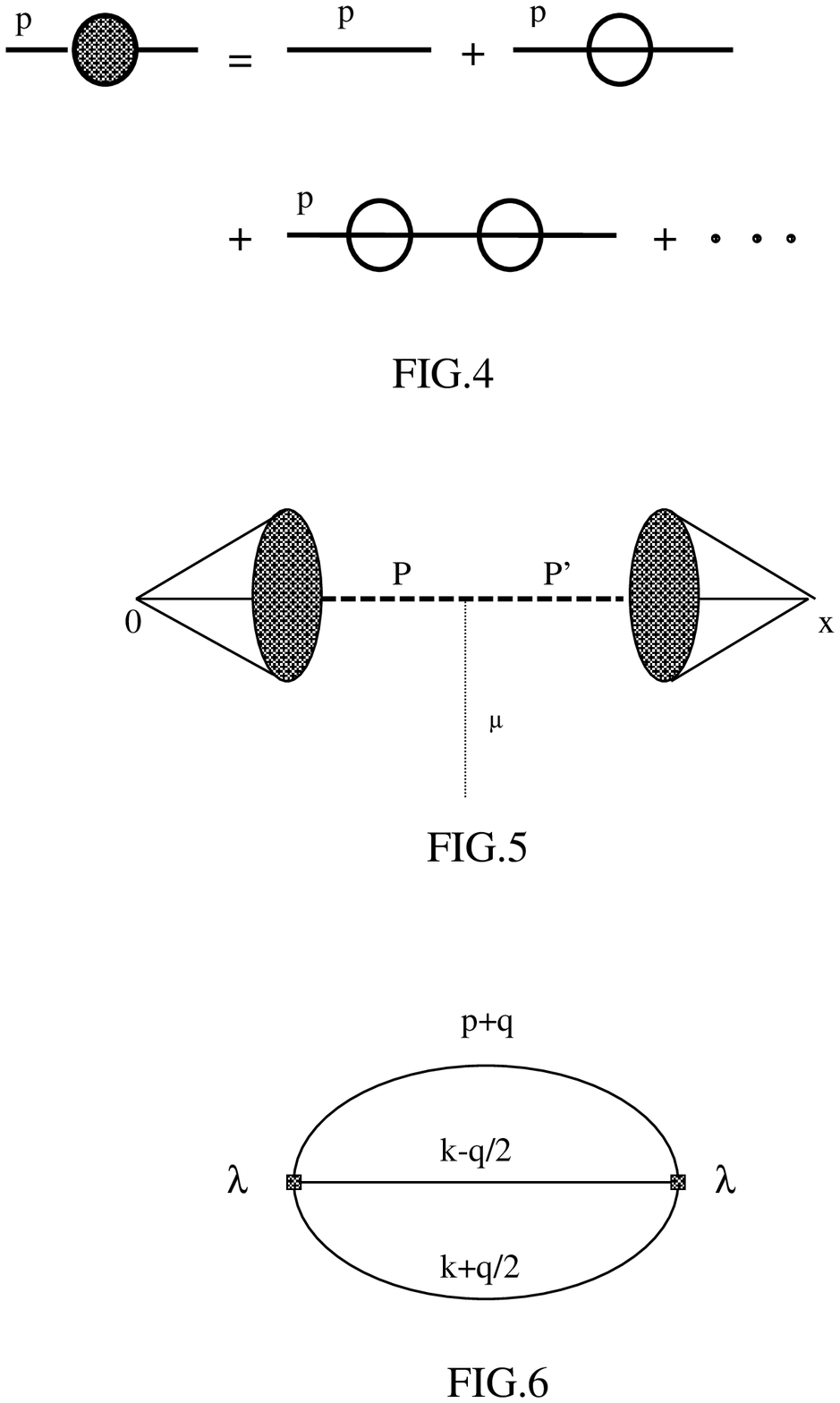}
}
\caption[]{Fig.5, Fig.6 and Fig.7}
\label{fig: fig4}
\end{figure}

\end{document}